\documentclass[aps,prb,twocolumn,superscriptaddress,howpacs]{revtex4}
\usepackage{graphicx}
\usepackage{color}
\usepackage{float}
\usepackage{bm}
\usepackage{braket}
\usepackage{placeins}
\usepackage{amsmath}

\begin{document}

\title{Acoustic plasmons and conducting carriers in hole-doped cuprate superconductors}

\author{A. Singh}
\author{H. Y. Huang}
\affiliation{National Synchrotron Radiation Research Center, Hsinchu 30076, Taiwan}

\author{Christopher Lane}
\affiliation{Theoretical Division, Los Alamos National Laboratory, Los Alamos, New Mexico 87545, USA}
\affiliation{Center for Integrated Nanotechnologies, Los Alamos National Laboratory, Los Alamos, New Mexico 87545, USA}

\author{J. H. Li}
\affiliation{Department of Physics, National Tsing Hua University, Hsinchu 30013, Taiwan}

\author{J. Okamoto}
\affiliation{National Synchrotron Radiation Research Center, Hsinchu 30076, Taiwan}

\author{S. Komiya}
\affiliation{Central Research Institute of Electric Power Industry, Yokosuka, Kanagawa, 240-0196, Japan}

\author{Robert S. Markiewicz}
\affiliation{Physics Department, Northeastern University, Boston, Massachusetts 02115, USA}

\author{Arun Bansil}
\affiliation{Physics Department, Northeastern University, Boston, Massachusetts 02115, USA}

\author{T. K. Lee}
\affiliation{Institute of Physics, Academia Sinica, Taipei 11529, Taiwan}
\affiliation{Department of Physics, National Sun Yat-sen University, Kaohsiung, 80424, Taiwan}
 
\author{A. Fujimori}
\affiliation{Department of Applied Physics, Waseda University, Shinjuku-ku, Tokyo 169-8555, Japan.}
\affiliation{National Synchrotron Radiation Research Center, Hsinchu 30076, Taiwan}

\author{C. T. Chen}
\affiliation{National Synchrotron Radiation Research Center, Hsinchu 30076, Taiwan}

\author{D. J. Huang}
\altaffiliation[Corresponding author:] {{
djhuang@nsrrc.org.tw}} 
\affiliation{National Synchrotron Radiation Research Center, Hsinchu 30076, Taiwan} \affiliation{Department of Physics, National Tsing Hua University, Hsinchu 30013, Taiwan}

\begin{abstract}
The layered crystal structures of cuprates enable collective charge excitations fundamentally different from those of three-dimensional metals. Acoustic plasmons have been observed in electron-doped cuprates by resonant inelastic X-ray scattering (RIXS); in contrast, whether acoustic plasmons exist in hole-doped cuprates is under debate, despite extensive measurements. This contrast led us to investigate the charge excitations of hole-doped cuprate La$_{2-x}$Sr$_x$CuO$_4$. Here we present incident-energy-dependent RIXS measurements and calculations of collective charge response via the loss function to reconcile the aforementioned issues. Our results provide evidence for the acoustic plasmons of Zhang-Rice singlet (ZRS), which has a character of the Cu $3d_{x^{2} - y^{2}}$ strongly hybridised with the O $2p$ orbitals; the metallic behaviour is implied to result from the movement of ZRS rather than the simple hopping of O $2p$ holes.
\end{abstract}

\date{\today}

\flushbottom
\maketitle

\thispagestyle{empty}

The superconductivity of cuprates, which has been a mystery ever since its discovery decades ago, is created through doping electrons or holes into a Mott insulator\cite{Keimer2015, Imada1998}.
There exists an inherent electron-hole asymmetry in cuprates\cite{segawa2010,Weber2010}. 
The antiferromagnetic phase of electron-doped cuprates persists at higher doping concentration and its superconductivity is more difficult  to achieve.  Electron correlations in electron-doped cuprates appear to be weaker than those of hole-doped compounds. Measurements of resonant inelastic X-ray scattering (RIXS) reveal acoustic plasmons in electron-doped cuprates\cite{Hepting2018,Lin2019},
showing the characteristic of conduction electrons of a layered system as shown  in Fig.~1(a)  in the presence of a long-range Coulomb interaction\cite{kresin1988, greco2019}. Acoustic plasmons of hole-doped cuprates, however,  were not observed in measurements of Cu $L$-edge RIXS\cite{Braicovich2010,LeTacon2011,Dean2013a,Minola2015,Minola2017,Ishii2017} and electron energy loss spectra (EELS)\cite{Mitrano2018,Husain2019}, despite that plasmons in cuprates have been observed  in other studies such as optical spectroscopy including reflectance and ellipsometry measurements\cite{Bozovic1990,Uchida1991,Marel2004,Levallois2016,Yin2019}, which is limited to nearly zero momentum transfer. In contrast, recent O $K$-edge RIXS results showed the presence of acoustic plasmons in hole-doped cuprates\cite{Nag2020}.

The conducting carriers are hallmarks of hole-doped cuprates, but their nature remains elusive. Upon doping, the low-energy excitations in the optical conductivity of La$_{2-x}$Sr$_x$CuO$_4$ (LSCO) are much enhanced\cite{Uchida1991};  the Hall coefficient $R_{\text{H}}$ of LSCO is much smaller than what is expected from the contribution of doped holes\cite{Ono2007}, i.e., $R_{\text{H}} \ll \frac{1}{{n_p}e}$, where $n_p$ is the hole density and $e$ is the magnitude of electron charge.
Early photoemission studies on superconducting cuprates indicated that the parent compound is a charge-transfer insulator, which led to a conjecture that doped holes are oxygen $p$-like \cite{fujimori1987}. Theoretically, doped holes are assumed to enter $p$ orbitals of an O atom between two Cu atoms, i.e., the Emery model \cite{emery1987}, or of four oxygen atoms surrounding one Cu atom to form a spin singlet termed a Zhang-Rice singlet (ZRS) \cite{ZRS1988}, as illustrated in Fig. 1(b). O $K$-edge X-ray absorption  \cite{chen1991} and  resonant elastic scattering  \cite{abbamonte2002} indeed provided experimental evidence for O $2p$ holes. The RIXS results also concluded that the acoustic plasmons are predominantly associated with the O sites through a strong $2p$ character\cite{Nag2020}. In contrast, photoemission intensity at the Fermi level showed no resonant enhancement at the O $K$-edge \cite{tjeng1992}. Many angle-resolved photoemission data \cite{Hashimoto2014} have been interpreted based on band structures in the local density approximation, which predict that Cu $d$ and oxygen $p$ characters are about equally mixed at the Fermi level. In addition, theoretical calculations using  the three-band Hubbard model\cite{CCChen2013} reveal that the ZRS band of hole-doped cuprates is composed of comparable amount of O $2p$ and Cu $3d_{x^{2} - y^{2}}$. 

 \begin{figure}[t]
\centering
\includegraphics[width=8.5cm]{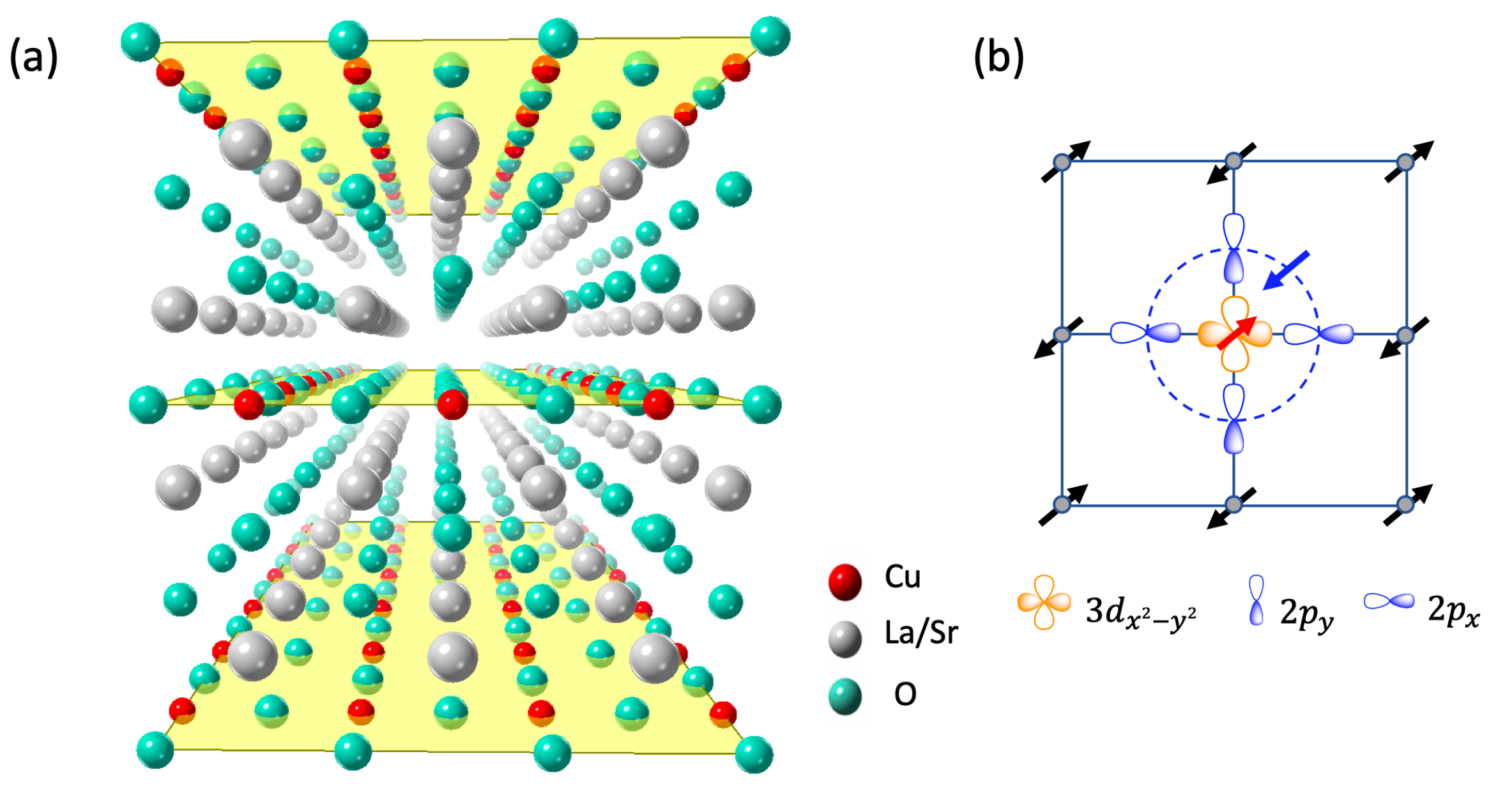}
\caption{(a) Crystal structure of LSCO. The stacked CuO$_2$ layers are highlighted in yellow. (b) Cartoon illustration of a ZRS in the CuO$_2$ plane. It is a two-hole state formed by a Cu $d_{x^{2} - y^{2}}$  hole hybridised with a O $2p_{x,y}$ hole distributed over the surrounding ligands in an antiparallel spin configuration. The singlet moves through the lattice of Cu$^{2+}$ ions in a way similar to that of a mobile hole in the in the CuO$_2$ plane. The  arrows on Cu$^{2+}$ indicate the antiferromagnetic structure of the lattice.}
\end{figure}

Here we report measurements of incident-energy-dependent O $K$-edge RIXS to investigate the nature of the conducting carriers of  superconducting LSCO with the doping concentration $x = 0.12$. Figure 2(a) shows the scattering geometry of our RIXS measurements. LSCO has a quasi-two-dimensional (2D) crystal structure with stacked CuO$_2$ layers . This system is theoretically expected to exhibit acoustic plasmons\cite{greco2019}.  Whereas optical  spectroscopy is limited to measurements of nearly zero momentum transfer, RIXS can probe the dispersion of acoustic plasmons in cuprates\cite{Hepting2018,Lin2019,Nag2020}. However, the conclusion drawn from O $K$-edge RIXS results\cite{Nag2020} are inconsistent with  EELS results\cite{Mitrano2018,Husain2019}. Our RIXS measurements explain the discrepancy between the EELS and RIXS results.

The parent compound La$_{2}$CuO$_4$ is an antiferromagnet, in which the Cu ion has an electronic configuration $3d^9$ with a hole of $d_{x^{2} - y^{2}}$ symmetry; the strong correlation effects split the conduction band into the lower and upper Hubbard bands (UHB). In the hole-doped phase, antiferromagnetic correlations are decreased and a metallic state appears, as demonstrated in  studies of the band theory\cite{Furness2018}.  The X-ray absorption spectrum\cite{chen1992}  (XAS) of O $K$-edge shows that the doped holes are on the O $2p_{x,y}$ orbitals hybridised with Cu $3d^{9}$. In the viewpoint of  the three-band Hubbard model, the O $2p$ hole denoted $\underline{L}$ and Cu $3d^{9}$ orbitals form a ZRS, i.e., $3d^{9}\underline{L}$, which plays the role of the lower Hubbard band of the one-band Hubbard or the $t-J$ model.  In the XAS of La$_{2}$CuO$_4$,  there exists a “prepeak” at the absorption edge arising from the UHB hybridised with the O $2p$ band. In the hole-doped compounds, a lower-energy absorption feature labelled ZRS emerges near 528.5 eV, as shown in the inset of Fig.~2(b). In the under-doped regime, this ZRS intensity increases linearly with hole doping \cite{chen1991}; hole doping thus manifests itself in the spectral weight transfer from UHB to ZRS as a consequence of electron correlations. For the O $1s$-to-$2p$ resonance,  an electron is excited from the $1s$ core level to ZRS or UHB; O $K$-edge RIXS spectra hence offer unique opportunities to probe the charge dynamics of hole-doped cuprates.

\begin{figure*}[ht]
\centering
\includegraphics[width=15cm]{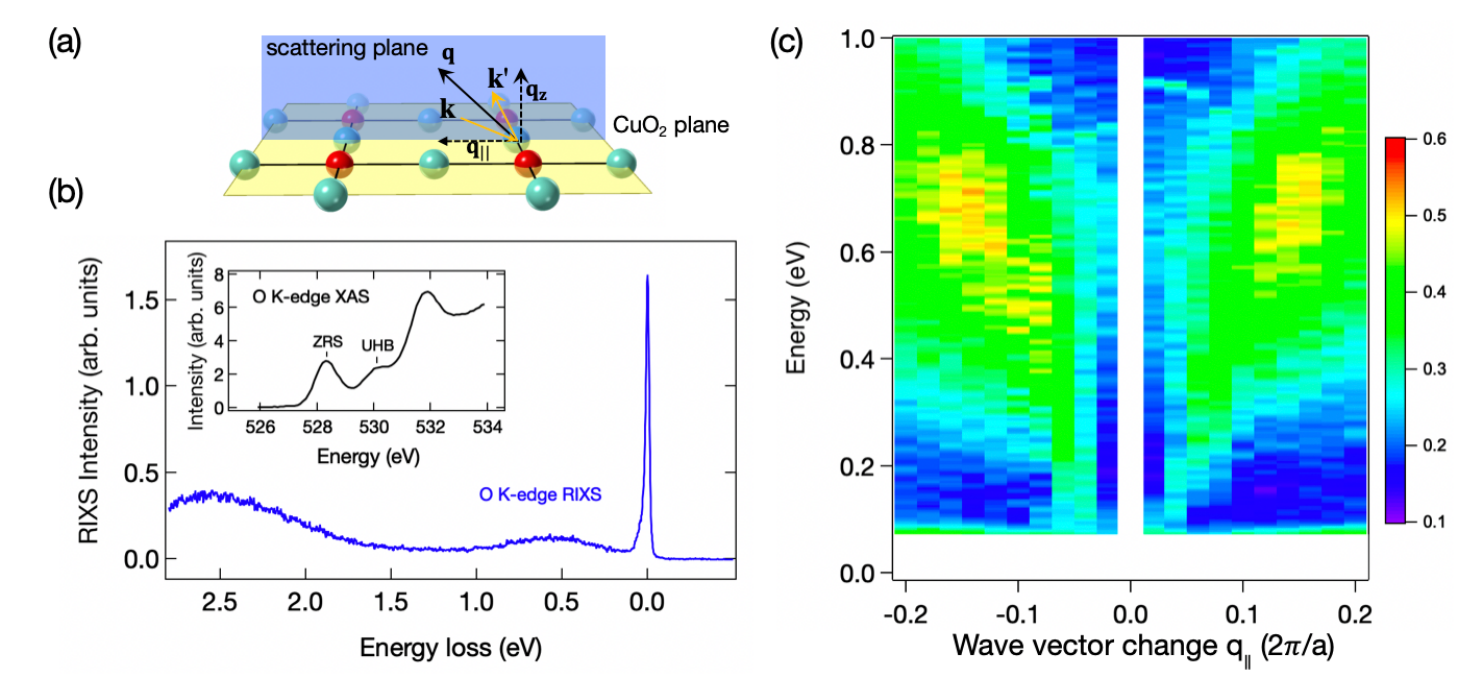}
\caption{(a) Scattering geometry of RIXS measurements. The CuO$_2$ plane is perpendicular to the scattering plane defined by \textbf{k} and  \textbf{k}$'$, which are the wave vectors of incident and scattered X-rays, respectively.  The projection of the wave vector change onto the CuO$_2$ plane, \textbf{q}$_\|$, is parallel to the antinodal direction.  (b) O $K$-edge RIXS spectrum of LSCO. This spectrum was recorded for a sample at temperature 23~K and with \textbf{q}$_{\|} = (0.1, 0)$ and $q_{z}=0.7$. The total energy resolution of the RIXS monochromator and spectrometer was 20~meV. The incident photon energy was set to the absorption at the ZRS resonance. Inset: XAS spectrum recorded through a fluorescence yield scheme. (c) RIXS intensity distribution map of LSCO in the energy-momentum space. The wave-vector change \textbf{q} is decomposed into an in-plane wave-vector change \textbf{q}$_{\|}$ that varies along the antinodal direction and $q_{z}$ is fixed to 0.7. RIXS spectra were recorded with incident X-rays tuned to the ZRS resonance.}
\end{figure*}

Figure~2(b) plots a typical RIXS spectrum of hole-doped cuprate LSCO with the incident photon energy set to the ZRS resonance in XAS. This spectrum includes the following features: elastic scattering, phonon excitations, plasmon and para-bimagnon excitations, $d$-$d$ excitations and electron-hole pair excitations.  Because of the limited energy resolution of the RIXS spectrometer, the phonon excitations were not well resolved from the elastic scattering, giving rise to an asymmetric and intense spectral profile near the zero energy loss. Because the O $2p$ bands are strongly hybridised with Cu $3d$, $d$-$d$ excitations of Cu and the excitonic excitations of ZRS occur in this RIXS spectrum at an energy loss above 1.5~eV; these excitation energies overlap with the energy of resonant fluorescence. Distinct from the Cu $L_3$-edge RIXS measurements, the O $K$-edge resonance yields negligible single-magnon excitations but permits us to observe even-order spin excitations. The excitation of para-bimagnons, which are called bimagnons throughout the paper for simplicity, appears in the broad feature at an energy loss centred at about 0.5 eV in LSCO\cite{Bisogni2012b,Chaix2018}.  This broad feature might also contain plasmon excitations, depending on the wave-vector change, which is composed of   in-plane wave-vector change \textbf{q}$_{\|}$  and out-of-plane component $q_z$, i.e.,  \textbf{q}~=~\textbf{q}$_{\|}~+~q_{z}{\bf {\hat z}}$; they are expressed in units of $\frac{2\pi}{a}$ and $\frac{2\pi}{c}$ throughout the paper,  respectively. Here $a$ and $c$ are the in- and out-of-plane lattice constants, respectively, and $d = c/2$ is the distance between two adjacent CuO$_2$ layers. 

We measured  \textbf{q}-dependent spectra on LSCO ($x=0.12$) to examine whether acoustic plasmons exist in hole-doped cuprates. Figure 2(c) maps the  distribution of RIXS intensity as a function of in-plane wave-vector change \textbf{q}$_{\|}$ along the antinodal direction with $q_z$ fixed to 0.7. This measurement method differs from those used in most of previous momentum-resolved RIXS measurements on cuprates in which $q_z$ was not fixed \cite{Lin2019,Ishii2017}. In addition to the quasi-elastic scattering, overall this RIXS intensity map shows a broad feature that shifts toward a higher energy with an increasing width as $q_{\|}$ is increased. After scrutinizing the RIXS data with \textbf{q}$_{\|}$ near the zone centre, we found that the RIXS spectrum of \textbf{q}$_{\|} = (0.04, 0)$ contains a low-energy narrow feature near 0.14~eV and a broad feature of bimagnon excitation centred at about 0.5 eV, like bimagnons in the undoped compound La$_2$CuO$_4$.

In the RIXS spectra recorded at the ZRS resonance, the bimagnon energy overlaps closely with the plasmon energy of $q_{\|}$  larger than 0.06. We resorted to curve fitting to extract the plasmon spectral weight from the contribution of the bimagnon. Each plasmon, bimagnon, and phonon spectral component was fitted with an anti-symmetrized Lorentzian function. The fitting also included one Gaussian function for elastic scattering. First, we fitted the spectrum of in-plane momentum $q_{\|}$ = 0.02 to extract the information about the energy and width of the bimagnon because its energy is well separated from the plasmon energy. For higher $q_{\|}$, several constraints on the energy and the width of bimagnons are required to achieve a reasonable fit so as to obtain the dispersion of the plasmon. See Supplementary Information for the details of curve fitting and fitted spectra. Figure 3(a) plots the spectra of various $q_{\|}$ after subtraction of the bimagnon and background; it reveals that the energy of the acoustic plasmon monotonically increases from 0.08 eV to 0.69 eV as  $q_{\|}$ is increased to 0.2.  The obtained plasmon dispersion of LSCO is similar to that of electron-doped cuprate La$_{2-x}$Ce$_x$CuO$_4$ (LCCO)\cite{Hepting2018}.  Our results indicate that the plasmons of hole-doped cuprates are strongly damped and have a greater spectral width than those of electron-doped cuprates, because of stronger electron correlations\cite{Kyung2004,Weber2010}. For example, the plasmon FWHM of LSCO at $q_{\parallel}= 0.1 $ and $q_{z}= 1.0$ is 0.68~eV, whereas that of LCCO is about 0.3~eV.

\begin{figure*}[t]
\centering
\includegraphics[width=16cm]{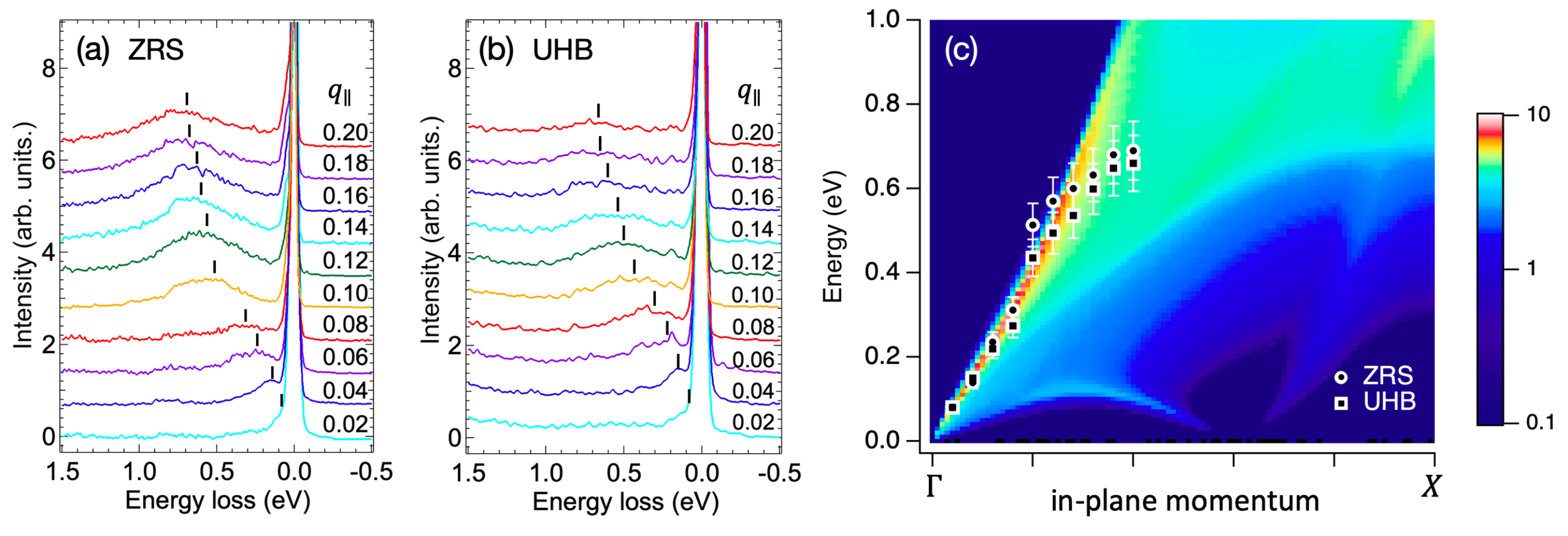}
\caption{ (a) \& (b) Momentum-dependent RIXS spectra of LSCO excited by X-rays of energy tuned to the ZRS and the UHB resonances, respectively. The out-of-plane component $q_{z}$ was fixed to 0.7.  Colour curves in (a) are spectra deduced from Fig.~2(c) after data smoothing and subtraction of the bimagnon component and the background. RIXS spectra shown in (b), which are multiplied by 3 to account for the difference in the cross section between the ZRS and UHB resonances, are plotted without bimagnon and background subtractions. The details of data analysis are presented in Supplementary Information.  Spectra in both (a) and (b) are offset vertically for clarity; vertical ticks indicate the plasmon energies.  (c) Theoretically obtained loss function for LSCO with hole-doping $x=0.12$ along $\Gamma-X$ in the square Brillouin zone with $q_{z}=0.7$. Measured plasmon energies of LSCO (circles \& squares ) deduced from the RIXS spectra shown in (a) and  (b) are also plotted for comparison. }
\end{figure*} 

The constraints of the bimagnon energy and width in curve fitting may cause some uncertainty about the plasmon dispersion, although the deduced dispersion is similar to that of electron-doped cuprates.  The conclusion from RIXS at the ZRS resonance requires further verification because of the spectral contributions of bimagnons. We further varied the incident photon energy in the RIXS measurements to diminish the spectral weight of the bimagnon. Figure 3(b) plots momentum-dependent RIXS spectra measured with the X-ray energy set to the UHB resonance, in which the excitations are dominated by Cu $3d$ orbitals.  The bimagnon contribution to the RIXS spectra in Fig. 3(b) is negligible, particularly for large $q_{\|}$ (Ref. \onlinecite{Bisogni2012b}). Magnons in hole-doped cuprates are created or removed when the ZRS hops between the two magnetic sublattices. As the UHB is more localised and about 2~eV above the ZRS band, RIXS excitations directly involved with the ZRS resonance are more sensitive to bimagnons than those with the UHB\cite{Bisogni2012b}. Figures 3(a) and 3(b)  show that the plasmon dispersions excited by X-rays tuned to the ZRS and UHB resonances agree with each other, revealing that both O $2p$ and Cu $3d$ orbitals are involved with the acoustic plasmons.

To verify that the observed collective charge dynamics follow the energy dispersion associated with acoustic plasmons, we calculated the collective charge response of the system via the loss function. That is, on incorporating a long-range Coulomb interaction into the one-band parameterization of LSCO (Ref. \onlinecite{markiewicz2008}) with a hole-doping, we calculated $\varepsilon^{-1}({\mathbf q},\omega)$, in which $\varepsilon$ is the dielectric function and \textbf{q} contains both in- and out-of-plane momentum transfer components, within the random phase approximation  (see Supplementary Information.). The three-dimensionality was incorporated through inter-plane Coulomb interactions. The loss function was eventually obtained as $-\text{Im}(\varepsilon^{-1})$. The calculations show that, for $q_{z} = 0$,  a clear plasmon peak of energy 0.7 eV appears at the zone center (see Supplementary Fig. S4),  in agreement with optical measurements\cite{Uchida1991}.

Figure~3(c) shows the calculated loss function for 12\% hole-doped LSCO in the paramagnetic phase along the $x$-axis in the Brillouin zone with $q_{z}=0.7$. The measured plasmon energies of LSCO are also shown for comparison. 
The comparison between calculated loss functions and the measured RIXS spectra for various  $q_{\parallel}$ is presented in Supplementary Fig. S7. 
The dispersion of an acoustic-like plasmon excitation is seen to extend from near zero at the zone centre $\Gamma$ to above 1 eV at $X$. The slight gap at $q_{\parallel}=(0,0)$ is due to the finite interlayer momentum transfer\cite{markiewicz2008,greco2016} in accord with the observed dispersion.
The charge fluctuations near the Fermi surface govern the plasmon excitations of small $q_{\|}$.
As $q_{\parallel}$ extends away from the zone centre, more incoherent states with decreased lifetime contribute to plasmon excitations and the width of the plasmon peak increases, consistent with the Landau quasi-particle picture in which the plasmon peak becomes incoherent. As the plasmon enters the particle-hole continuum the peak broadens and the slight curve appears to follow the ridge in the particle-hole continuum of the loss function. 
Consequently, the RIXS spectral width of acoustic plasmons becomes broadened when $q_{\|}$ increases, consistent with the dynamics of the electrons near the Fermi surface. The agreement between our data and theory thus supports the proposal of attributing the zone centre mode to acoustic plasmon excitations.

Next, we corroborate the three dimensional (3D) nature of an acoustic plasmon originating from the interlayer Coulomb interaction. Figures 4(a) plots the spectra of \textbf{q}$_{\|} = (0.1, 0)$ with selected $q_z$. The broad spectral features about 0.5~eV were fitted with two anti-symmetrized Lorentzian functions: one is a $q_z$-independent component for the bimagnon; the other disperses with the change of  $q_z$. See Supplementary Information for the details of data analysis.   We found that, when $q_z$ is altered from 0.6 to 1.0, the energy of this feature decreases  by 91~meV for $q_{\|}$ fixed to 0.1, i.e., evidence for the 3D nature of plasmon excitations. Figure 4(b) illustrates the dispersions of the plasmon bands of a layered electron-gas model for the high-$T_{\rm c}$ cuprates\cite{kresin1988}. The plasmon bands are restricted to be in between two boundary branches of $q_z=0$ and $q_z=\frac{\pi}{d}$, which correspond to the in-phase motion of electrons on separate planes and the out-of-phase motion on adjacent planes, respectively. Figure 4(c) compares the measured acoustic-plasmon energy with that from the calculated loss function plotted in Fig. S4 for $q_{\|} = 0.1$. The calculations further verify the decreased plasmon energy as being due to the increase of $q_z$ and reveal the plasmon origin of the observed RIXS excitation.

In combination with the absence of acoustic plasmon in Cu $L$-edge RIXS measurements\cite{Braicovich2010,LeTacon2011,Dean2013a,Minola2015,Minola2017,Ishii2017}, the results of O $K$-edge RIXS tuned only to the ZRS resonance seem to suggest that the acoustic plasmons are predominantly of a O $2p$ character\cite{Nag2020}. This conclusion, however, is in contrast to the ZRS picture\cite{ZRS1988}, in which O $2p$ and Cu $3d$ are strongly hybridised and the singlet hops through the Cu$^{2+}$ lattice in a way similar to a hole in the single-band effective Hamiltonian. In hole-doped cuprates, the ZRS band crosses the Fermi level at wave vectors near $(\pi, 0)$ and $(\tfrac{\pi}{2}, \tfrac{\pi}{2})$, as shown in angle-resolved photoemission measurements\cite{Yoshida2003, Shen2004}. 
One can use a simple cluster model\cite{Eskes1990} to comprehend the spectral weight of the transitions from the ground state to the unoccupied ZRS and UHB probed by O $K$-edge RIXS; the former has a comparable spectral weight of O $2p$ and Cu $3d$ orbitals, whereas the latter is dominated by Cu $3d$. See the discussion in Supplementary Information.  Calculations using the three-orbital Hubbard model\cite{CCChen2013} for hole-doped cuprates also indicate that  the spectral weight  of Cu $3d_{x^{2} - y^{2}}$  in the ZRS band  is comparable to that of O $2p$, and that the UHB is dominated by Cu $3d_{x^{2} - y^{2}}$. Our observation of plasmons at both ZRS and UHB resonances unravels that the acoustic plasmons are of ZRS character, consistent with the delocalized nature of the ZRS.  In other words, not only O $2p$ but also Cu $3d$ states contribute to the intinerant motion of charge carriers in hole-doped cuprates. 

\begin{figure*}[t]
\centering
\includegraphics[width=16cm]{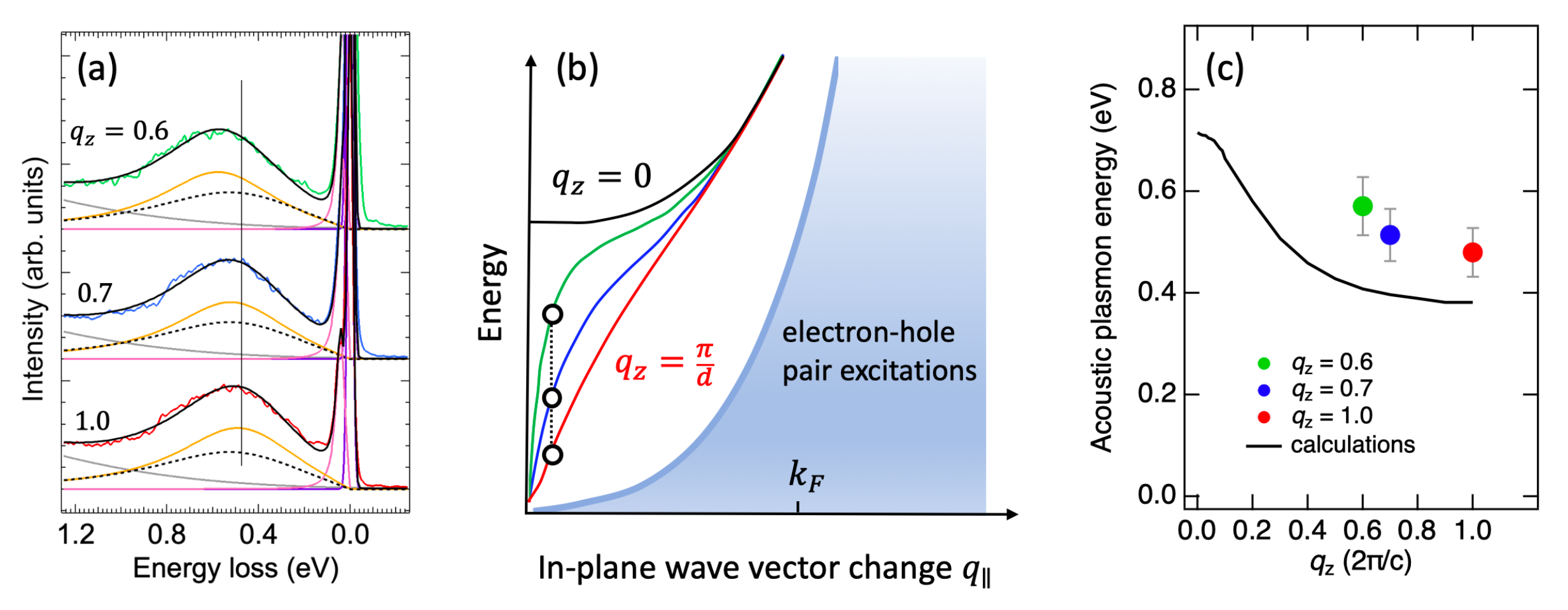}
\caption{3D nature of acoustic plasmons. (a) RIXS spectra of LSCO after data smoothing for selected $q_z$ with $q_{\|}$ fixed to 0.1. The incident X-ray energy was tuned to the ZRS resonance. The fitted spectral function for the plasmon and bimagnon is the anti-symmetrized Lorentzian function, as plotted in orange curves and black dashed lines, respectively; the background curves are plotted in grey lines. Bimagnon spectra of all $q_{z}$ have the same energy position and width. Spectra are vertically offset for clarity. The vertical line indicates the plasmon energy of $q_{z}=1.0$. See Supplementary Information for the details of curve fitting.
(b) Graphic illustration of the dispersion relations of selected plasmon bands and the electron-hole pair excitations of a layered electron-gas model (after Ref. \onlinecite{kresin1988}). The circles indicate the plasmon energy in response to the change in $q_z$ with $q_{\|}$ fixed. (c) Comparison of the measured acoustic plasmon energies with those from the calculated loss function for $q_{\|} = 0.1$. The measured plasmon energies plotted as coloured circles correspond to spectra shown in (a) from which the plasmon energies were deduced through curve fitting. The plasmon energy plotted by the black line is from the calculated loss function shown in Supplementary Information.}
\end{figure*} 

In conclusion, our observation of acoustic plasmons sheds light on the nature of the conducting carriers of hole-doped cuprates. O $K$-edge RIXS results demonstrate that the $p$ and $d$ orbitals are hybridised in ZRS bands and UHB, consistent with XAS results. The observation of plasmons at both UHB and ZRS resonances corroborates the ZRS picture, revealing that the acoustic plasmons indeed result from the delocalized nature of the ZRS. Our findings indicate that the conducting carriers are of ZRS rather than O $2p$ holes.

\vspace{1cm}
\noindent {\bf{ Methods}}
\vspace{0.1cm}

The LSCO single crystal with $x=0.12$ was grown by the traveling-solvent floating zone method\cite{Komiya2002, Komiya2005, Ono2007}. After growth, the crystals were annealed appropriately to remove oxygen defects. The value of $x$ was determined from an inductively-coupled-plasma atomic-emission spectrometric analysis. The LSCO single crystal has lattice constants $a = b=3.78$~{\AA} and $c = 13.2$~{\AA}.

O $K$-edge RIXS measurements were conducted at a newly constructed AGM-AGS spectrometer\cite{Singh2021} of  beamline 41A at Taiwan Photon Source in National Synchrotron Radiation Research Center (NSRRC), Taiwan.  RIXS spectra were recorded with $\sigma$-polarized incident X-rays; the sample was at 23~K.  Before the RIXS measurements, a clean sample surface (001) was obtained on cleaving the sample in air.  The instrumental energy resolution was ${\Delta}E \approx$ 20~meV FWHM for all RIXS measurements.  All RIXS spectra were normalized to the fluorescence intensity. See Supplementary Information for the experimental setup.

We calculated the loss function using the random phase approximation and the one-band tight-binding model with the full long-range Coulomb interaction carefully taken into account\cite{markiewicz2007, markiewicz2008}. See Supplementary Information for the details.

\vspace{1cm}
\noindent {\bf{ Acknowledgements}}
\vspace{0.1cm}

\noindent 

We acknowledge NSRRC staff for technical support. We thank Teppei Yoshida for technical advice on cleaving LSCO crystals. 
This work was supported in part by the Ministry of Science and Technology of Taiwan under Grant No. 103-2112-M-213-008-MY3 and MOST 108-2923-M-213-001.
We acknowledge the support by Japan Society for the Promotion of Science under Grant No. 19K03741.
The work at Northeastern University was supported by the US Department of Energy (DOE), Office of Science, Basic Energy Sciences grant number DE-FG02-07ER46352 and benefited from Northeastern University's Advanced Scientific Computation Center (ASCC), and the NERSC supercomputing centre through DOE grant number DE-AC02-05CH11231. The work at Los Alamos National Laboratory was supported by the U.S. DOE NNSA under Contract No. 89233218CNA000001 and by the Center for Integrated Nanotechnologies, a DOE BES user facility, in partnership with the LANL Institutional Computing Program for computational resources. Additional support was provided by DOE Office of Basic Energy Sciences Program E3B5.

\bibliographystyle{naturemag}
\bibliography{ms.bib}
\end{document}


\preprint{APS/123-QED}

\title{Supplementary Information \--- Acoustic plasmons and conducting carriers in hole-doped cuprate superconductors}

\author{A. Singh}
\author{H. Y. Huang}
\affiliation{National Synchrotron Radiation Research Center, Hsinchu 30076, Taiwan}

\author{Christopher Lane}
\affiliation{Theoretical Division, Los Alamos National Laboratory, Los Alamos, New Mexico 87545, USA}
\affiliation{Center for Integrated Nanotechnologies, Los Alamos National Laboratory, Los Alamos, New Mexico 87545, USA}

\author{J. H. Li}
\affiliation{Department of Physics, National Tsing Hua University, Hsinchu 30013, Taiwan}

\author{J. Okamoto}
\affiliation{National Synchrotron Radiation Research Center, Hsinchu 30076, Taiwan}

\author{S. Komiya}
\affiliation{Central Research Institute of Electric Power Industry, Yokosuka, Kanagawa, 240-0196, Japan}

\author{Robert S. Markiewicz}
\affiliation{Physics Department, Northeastern University, Boston, Massachusetts 02115, USA}

\author{Arun Bansil}
\affiliation{Physics Department, Northeastern University, Boston, Massachusetts 02115, USA}

\author{T. K. Lee}
\affiliation{Institute of Physics, Academia Sinica, Taipei 11529, Taiwan}
\affiliation{Department of Physics, National Sun Yat-sen University, Kaohsiung, 80424, Taiwan}

\author{A. Fujimori}
\affiliation{Department of Applied Physics, Waseda University, Shinjuku-ku, Tokyo 169-8555, Japan.}
\affiliation{National Synchrotron Radiation Research Center, Hsinchu 30076, Taiwan}

\author{C. T. Chen}
\affiliation{National Synchrotron Radiation Research Center, Hsinchu 30076, Taiwan}

\author{D. J. Huang}
\altaffiliation[Corresponding author:] {{
djhuang@nsrrc.org.tw}} 
\affiliation{National Synchrotron Radiation Research Center, Hsinchu 30076, Taiwan} 
\affiliation{Department of Physics, National Tsing Hua University, Hsinchu 30013, Taiwan}

\date{\today}

\maketitle

\subsection{O $K$-edge RIXS Measurements}
O $K$-edge resonant inelastic X-ray scattering (RIXS) measurements were conducted at the AGM-AGS spectrometer of beamline 41A at Taiwan Photon Source in National Synchrotron Radiation Research Center. This AGM-AGS beamline has been recently constructed based on the energy compensation principle of grating dispersion. The instrument has a best energy resolution of 16 meV at full width half maximum (FWHM) at 530 eV photon energy. For a routine RIXS measurement, the spectrometer angle 2$\theta$ can swing a wide angle from 65$^\circ$ to 150$^\circ$. The total energy resolution was  ${\Delta}E$ = 20 meV FWHM for all RIXS measurements with the monochromator exit slit set to 100 ${\mu}$m. 
For more details of the RIXS beamline see Ref. \cite{Singh2021}.

The La$_{2-x}$Sr$_x$CuO$_4$ (LSCO) sample was cleaved in air and then mounted on the 3-axis in-vacuum manipulator through a loadlock system. Prior to RIXS measurements, the crystallographic axes were precisely aligned with hard X-ray diffraction using a special holder with tilting adjustment. X-ray absorption spectrum was measured using a photodiode in the fluorescence yield mode. The resonant conditions were achieved by tuning the energy of the incident X-ray to tthe Zhang-Rice singlet (ZRS) resonance near O $\textit{K}$-edge, about 528.3 eV. The sample was cooled to 23 K with liquid He. RIXS measurements were carried out as a function of in-plane wave-vector change ${\bf q}_{\|} = (q_{\|}, 0)$ ${2\pi/a}$ with $q_{z}$ fixed at 0.6, 0.7, or 1.0 ${\pi/d}$, where $d$ is the distance between two adjacent CuO$_2$ planes. ${\bf q}_{\|}$ was varied from $(-0.2, 0)$ ${2\pi/a}$ to $(0.2, 0)$ ${2\pi/a}$ in steps of 0.01 and RIXS spectra were measured for each ${\bf q}_{\|}$ with 1 hour exposure time.
RIXS measurements with the incident X-ray energy set to ZRS or UHB provide us with an opportunity to  differentiate the contributions of $p$ holes to the acoustic plasmons from those contributed by $d$ holes. We also measured RIXS spectra with the X-ray absorption energy set to UHB (529.9 eV) for ${\bf q}_{\|}$ from $(0, 0)$ to $(0.2, 0)$ with $q_{z}$ fixed to 0.7 ${\pi/d}$. $q_{z}$-dependent measurements were carried out with fixed ${\bf q}_{\|} = (0.04, 0)$ and ${\bf q}_{\|} = (0.1, 0)$. RIXS spectra for  $q_{z}$ = 0.6 ${\pi/d}$, 0.7 ${\pi/d}$ and 1.0 ${\pi/d}$ were measured for each ${\bf q}_{\|}$ with an exposure time of 4 hours.

\subsection{RIXS Data Analysis and Curve Fitting}
We analysed the RIXS spectra measured at ZRS and UHB resonances through a non-linear least-squares curve fitting. Prior to the fitting, spectra were normalized to the incident photon flux and corrected for geometry effect \cite{achkar2011bulk}. Each of the plasmon, bimagnon, and phonon components was fitted with an anti-symmetrized Lorentzian function $f(\omega)$ expressed as 

\begin{widetext}
\begin{align}
f(\omega)=\frac{\gamma}{2\pi}\left[\frac{1}{(\omega-\omega_0)^2+(\gamma/2)^2}-\frac{1}{(\omega+\omega_0)^2+(\gamma/2)^2}\right],
\label{eq_1}
\end{align}
\end{widetext}
where $\omega_0$ is the transition energy and $\gamma$ is FWHM \cite{Hepting2018, Nag2020}.
A cubic background was used to account for  the tail contribution of $d-d$ excitation. In addition, we used a Gaussian function for the component of elastic scattering.

Figure S1 shows the RIXS spectra measured with ZRS resonance energy along with the fitted components. 
First, we fitted the spectrum of in-plane momentum $q_{\|}$ = 0.02 with four components: one Gaussian function for elastic scattering with an instrumental energy resolution of 20 meV, and three anti-symmetrized Lorentzian functions for bimagnon, plasmon, and phonon of energy at $\approx$ 0.03 to 0.05 eV. We limited the bimagnon energy to be larger than 0.4 eV and found that it was a broad peak centered about 0.5 eV, consistent with published results \cite{Bisogni2012b, Nag2020}.
For the fitting of RIXS spectra of other in-plane momenta, the position and FWHM of the bimagnon component were constrained to be close to those obtained from the fit results of $q_{\|}$ = 0.02 within $\pm$0.025 eV and $\pm$0.15 eV, respectively. For the fitting of $q_z$-dependent data, the position, amplitude and width of the bimagnon component was fixed to the values obtained from in-plane momentum analysis. It is evident from Fig. S1 that for higher $q$ ($q_{\|}$ from 0.12 to $q_{\|}$ = 0.20) the bimagnon intensity decreases significantly.

Furthermore, we fitted the RIXS spectra measured at the UHB resonance with the same scheme used for the analysis of data measured at the ZRS resonance. Figure S2 shows those spectra after smoothing along with the fitted components. Clearly the bimagnon contribution to the RIXS spectra measured with UHB is nearly negligible, particularly for high $q_{\|}$. Since the UHB is more localised and at about 2 eV higher than the ZRS band, RIXS excitations directly involved with the ZRS resonance are more sensitive to bimagnons than
those with the UHB \cite{Bisogni2012b}. In addition a low energy exciton excitation peak $\approx 0.18-0.2$ eV is observed for high $q_{\|}$ as explained in Ref \cite{lane2020landscape}. Table S1 lists the fitted parameters of our curve fitting analysis. 

\begin{table}[h]
\caption{Fit results of RIXS spectra recorded at the ZRS and UHB resonances for various in-plan momentum ($q_{\|}$, 0) in units of $2\pi/a$. The energy of plasmon ($\omega_\textrm{plasmon}$) and bimagnon ($\omega_\textrm{bimag}$) are given in units of eV; their FWHM are expressed by $\gamma_\textrm{plasmon}$ and $\gamma_\textrm{bigmag}$ in units of eV, respectively.}
\begin{ruledtabular}
\begin{tabular}{lllllll}
&$q_{\|}$& $\omega_\textrm{plasmon}$ &
$I_\textrm{plasmon}$ &
$\gamma_\textrm{plasmon}$  & $\omega_\textrm{bimag}$ & $\gamma_\textrm{bimag}$ \\
\hline
ZRS\\
&0.02 & 0.08 & 0.18 & 0.23 & 0.55 & 0.78  \\
&0.04 & 0.14 & 0.34 & 0.26 & 0.54 & 0.75  \\
&0.06 & 0.24 & 0.35 & 0.43 & 0.55 & 0.77  \\
&0.08 & 0.31 & 0.30 & 0.56 & 0.50 & 0.79  \\
&0.10 & 0.51 & 0.67 & 0.65 & 0.50 & 0.90  \\
&0.12 & 0.57 & 1.10 & 0.68 & 0.50 & 0.80  \\
&0.14 & 0.60 & 1.13 & 0.70 & 0.50 & 0.80  \\
&0.16 & 0.63 & 1.15 & 0.72 & 0.50 & 0.90  \\
&0.18 & 0.68 & 1.03 & 0.73 & 0.50 & 0.63  \\
&0.20 & 0.69 & 0.99 & 0.76 & 0.50 & 0.64  \\
UHB\\
&0.02 & 0.08 & 0.07 & 0.18 & 0.50 & 0.61  \\
&0.04 & 0.15 & 0.09 & 0.20 & 0.50 & 0.61  \\
&0.06 & 0.22 & 0.13 & 0.35 & 0.50 & 0.60  \\
&0.08 & 0.30 & 0.24 & 0.60 & -- & --  \\
&0.10 & 0.44 & 0.22 & 0.62 & -- & --  \\
&0.12 & 0.49 & 0.24 & 0.63 & -- & --  \\
&0.14 & 0.54 & 0.22 & 0.65 & -- & --  \\
&0.16 & 0.60 & 0.22 & 0.67 & -- & --  \\
&0.18 & 0.65 & 0.20 & 0.69 & -- & --  \\
&0.20 & 0.66 & 0.18 & 0.73 & -- & --  \\
\end{tabular}
\end{ruledtabular}
\end{table}

\begin{figure*}[t]
\centering
\includegraphics[width=17cm]{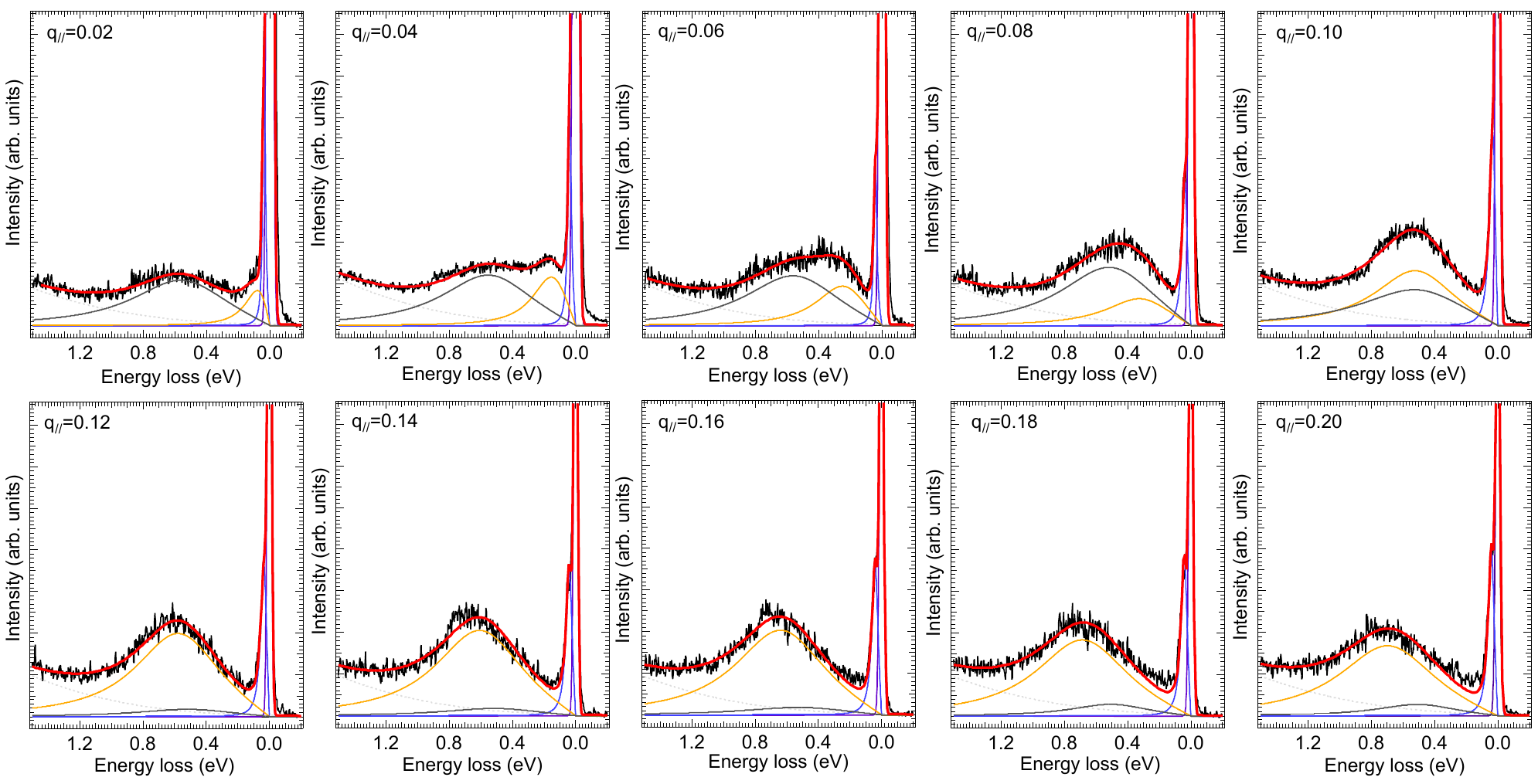}
\caption{Momentum dependent RIXS spectra of LSCO along with the fitted components. RIXS spectra were recorded with incident X-ray tuned to the ZRS resonance. Black solid curves represent the data; solid red lines plot the sum of the fitted components, which include a elastic peak (purple solid line), phonon excitation (blue solid line) and bimagnon (gray solid lines). Light gray dashed lines represent a cubic background.}
\end{figure*}

\begin{figure*}[!bh]
\centering
\includegraphics[width=17cm]{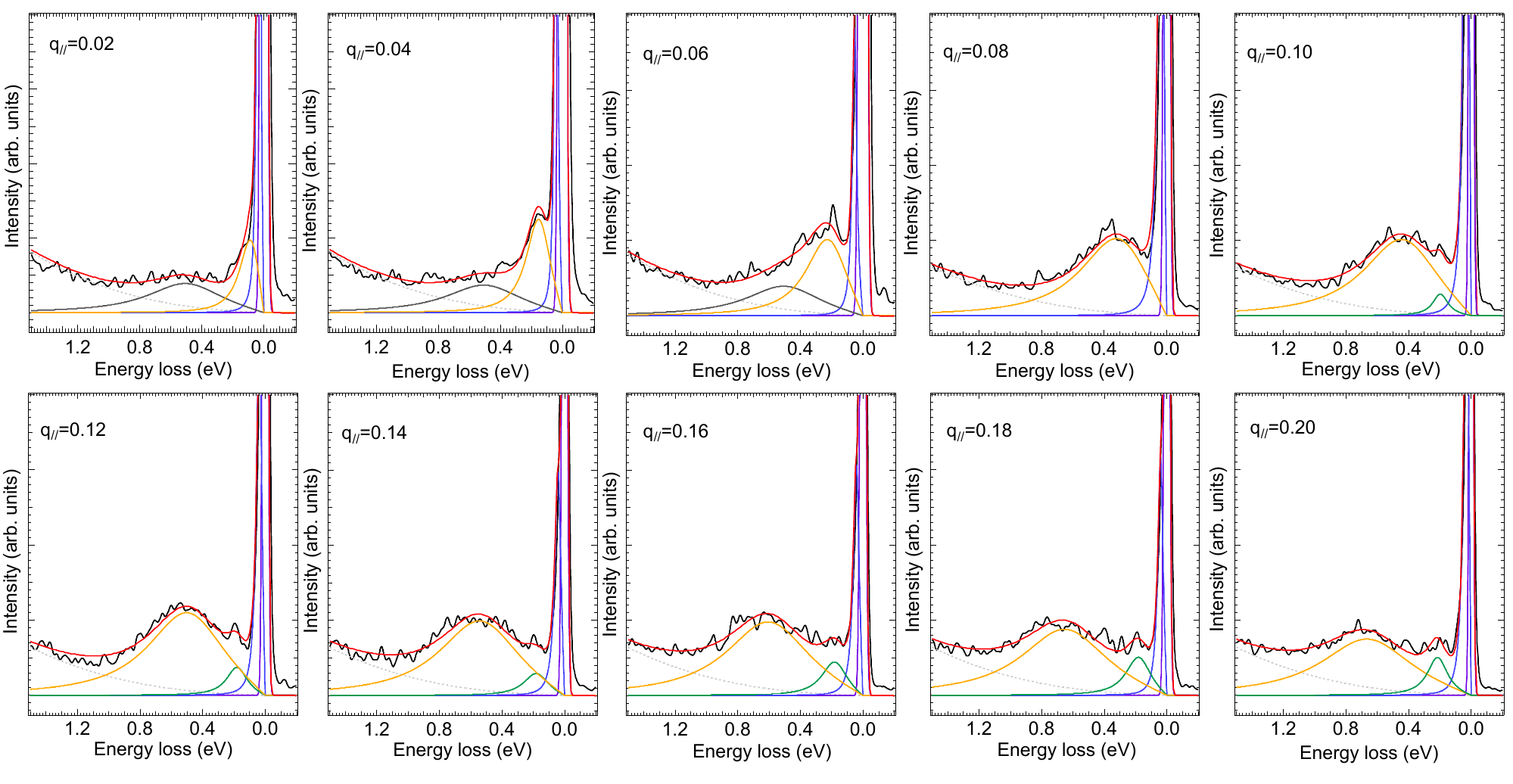}
\caption{Momentum dependent RIXS spectra of LSCO along with the fitted components. RIXS spectra were taken with incident X-ray tuned to the UHB resonance. Black solid curves represents the data after smoothing, and solid red lines plots the sum of the fitted components, which includes as elastic peak (purple solid line), phonon excitation (blue solid lines) and bimagnon (gray solid lines). For high $q_{//}$, a low energy exciton excitation is represented by green solid lines. Light gray dashed lines represent a cubic background.}
\end{figure*}

\begin{figure}[ht]
\centering
\includegraphics[width=8.5cm]{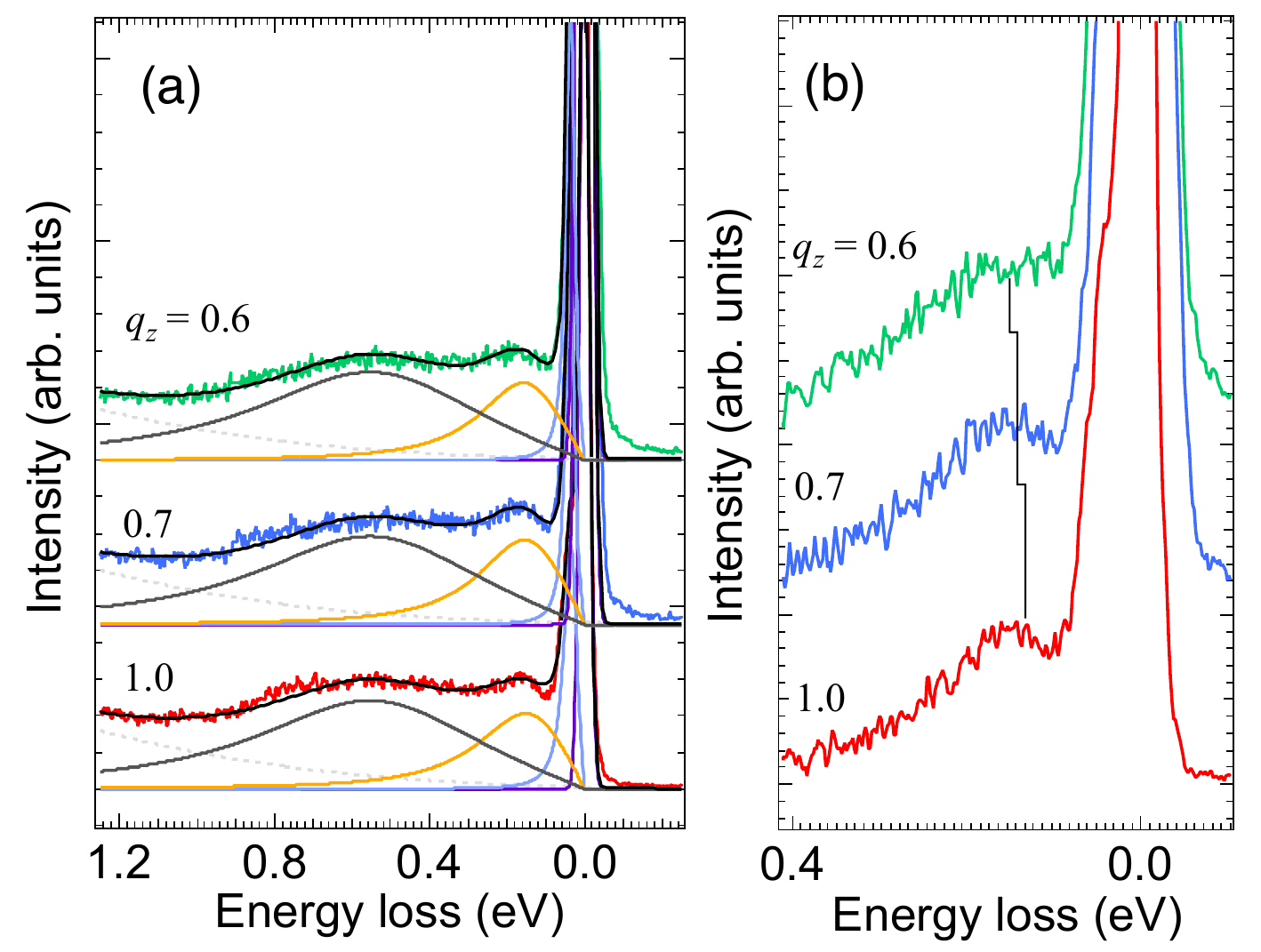}
\caption{(a) RIXS spectra of LSCO for selected $q_z$ with $q_{\|}$ fixed to 0.04. The incident X-ray energy was tuned to the ZRS resonance. The fit components of plasmon and bimagnon are the anti-symmetrized Lorentzian function, as plotted in orange and grey curves; the background are plotted by dotted  lines. Para-bimagnon spectra of all $q_{z}$ have the same energy position and width.
(b)RIXS spectra after subtraction of the bimagnon component and the background  from those shown in (a). The vertical line indicates the shift of acoustic plasmon energy obtained from curve fitting due to the change of $q_{z}$. Spectra in both (a) and (b) are vertically offset for clarity. }
\end{figure}

Figure S3(a) plots the spectra of \textbf{q}$_{\|} = (0.04, 0)$ with selected $q_z$, along with the fitted components. Through the curve fitting, we found that, when $q_z$ is changed from 0.6 to 1.0, the energy of this feature decreases  by 15~meV. Figure S3(b) plots the spectra after subtraction of the bimagnon component and the background from Fig. S3(a). The vertical line indicates the shift of acoustic plasmon energy obtained from curve fitting due to the change of $q_{z}$. These plots show that the excitation energy of the narrow feature disperses with the change of $q_z$. 

\subsection{Cluster Model for Analysis of RIXS Spectral Weight}

In O $K$-edge RIXS, the O $1s$ electron is excited to the ZRS or UHB. Theoretically one can use a model Hamiltonian and the Kramers-Heisenberg formula to calculate RIXS cross section \cite{Ament2011}. 
To comprehend the spectral weight  of a RIXS process, we approximate the excitation as a transition from the ground state to an excited state.
We use a simple cluster model \cite{Eskes1990} to obtain the spectral weight of the transitions from the ground state to the ZRS and UHB probed by O $K$-edge RIXS. 
In the O $K$-edge RIXS of cuprates, we neglect the effect of core hole in the intermediate state to obtain a reasonable approximation. For a cluster model of a Cu$^{2+}$ ion surrounded by four oxygens, the ground state is described as 
\begin{equation}
\ket{\rm{GS}} = \alpha \ket{d^9}+ \beta \ket{d^{10}\underline{L}}, 
\end{equation}
and the ZRS and UHB states are 
\begin{align}
&\ket{\rm{ZRS}} = \gamma \ket{d^8}+ \delta \ket{d^{9}\underline{L}}+\epsilon\ket{d^{10}\underline{L}^2}\\
&\ket{\rm{UHB}} = \ket{d^{10}},
\end{align}
where $\underline{L}$ represents a ligand hole. The RIXS cross section at the ZRS and UHB resonances are approximately proportional to  $|\bra{\rm{GS}}\hat{a}_{i}^{\dag}\ket{\rm{ZRS}}|^2$ and $\left|\bra{\rm{GS}}\hat{a}_{i}\ket{\rm{UHB}}\right|^2$ ($i$~=~Cu $3d$, O $2p$), respectively, where $\hat{a}_{i}^\dag$ ($\hat{a}_{i}$) is the electron creation (annihilation) operator on orbital $i$.
We extract the spectral weights of O $2p$ and Cu $3d$ at the ZRS resonance as follows:
\begin{equation}
I_{\scaleto{\braket{\rm{GS}|\rm{ZRS}}}{5pt}}^{\scaleto{O2p}{5pt}} = \left|\alpha\delta+\beta\epsilon\right|^2 
\end{equation}
and
\begin{equation}
I_{\scaleto{\braket{\rm{GS}|\rm{ZRS}}}{5pt}}^{\scaleto{Cu3d}{4pt}} = \left|\alpha\gamma+\beta\delta\right|^2.
\end{equation}
Similarly, the spectral weights of O $2p$ and Cu $3d$ at the UHB resonance are:
\begin{align}
&I_{\scaleto{\braket{\rm{GS}|\rm{UHB}}}{5pt}}^{\scaleto{O2p}{5pt}} = \left|\beta\right|^2\\ 
&I_{\scaleto{\braket{\rm{GS}|\rm{UHB}}}{5pt}}^{\scaleto{Cu3d}{4pt}} = \left|\alpha\right|^2.
\end{align}
From the coefficients given by Eskes et al. for CuO \cite{Eskes1990}, we have $\alpha=\sqrt{0.67}$, $\beta=\sqrt{0.33}$, $\gamma=\sqrt{0.07}$, $\delta=\sqrt{0.64}$, and 
$\epsilon=\sqrt{0.28}$.
This gives 
$I_{\scaleto{\braket{\rm{GS}|\rm{ZRS}}}{5pt}}^{\scaleto{O2p}{5pt}}=0.86$, $I_{\scaleto{\braket{\rm{GS}|\rm{ZRS}}}{5pt}}^{\scaleto{Cu3d}{4pt}}=0.43$, $I_{\scaleto{\braket{\rm{GS}|\rm{UHB}}}{5pt}}^{\scaleto{O2p}{5pt}}=0.33$, and $I_{\scaleto{\braket{\rm{GS}|\rm{UHB}}}{5pt}}^{\scaleto{Cu3d}{4pt}}=0.67$.
Therefore, O $2p$ and Cu $3d$ orbitals have a comparable spectral weight in the RIXS transition at the ZRS resonance, whereas the transition to the UHB is dominated by Cu $3d$. (Note: That $I_{\scaleto{\braket{\rm{GS}|\rm{ZRS}}}{5pt}}^{\scaleto{O2p}{5pt}}+ I_{\scaleto{\braket{\rm{GS}|\rm{ZRS}}}{5pt}}^{\scaleto{Cu3d}{4pt}}$  exceeds 1 is due to the so-called "dynamical effect" of spectral weight transfer widely seen in correlated systems.)

\subsection{Theoretical Calculations of the Loss Function}
In a RIXS process, a photon scatters resonantly to another
state, leaving behind an electron-hole excitation of well
defined momentum $\mathbf{q}$ and energy $\omega$. The general expression for the $K$-edge RIXS intensity\cite{nomura2005analysis,markiewicz2006collapse} can be expressed as\cite{markiewicz2008}
\begin{align}
I_{RIXS}(\mathbf{q},\omega,\omega_i)=(2\pi)^3 N\left|w(\omega,\omega_i)\right|^2 \frac{\hat{v}(\mathbf{q})^2}{v(\mathbf{q})}L(\mathbf{q},\omega)
\end{align}
where $w(\omega,\omega_i)$ is the energy dependent matrix element, $\hat{v}(\mathbf{q})$ ($v(\mathbf{q})$) is the interaction energy of between electron-hole (electron-electron), and $L(\mathbf{q},\omega)$ is the loss function    in the charge channel 
\begin{align}
\text{Im}\left[-\varepsilon^{-1}_{00}\right],
\end{align}
with $\varepsilon^{-1}_{00}$ being the inverse dielectric function. Therefore, peaks $K$-edge RIXS intensity are directly poportional to transitions in the loss function, which mark the presence of collective charge modes, or plasmons \cite{Sturm2013}. To construct the inverse dielectric function we follow Ref. \onlinecite{Aryasetiawan2008}, where it is defined as
\begin{align}\label{eq:dielectric}
\varepsilon^{-1}_{IJ}=\left[ \delta_{NL}-v_{NK}P_{KL} \right]^{-1}_{IJ}.
\end{align}
$I(J)=0$ index the charge components, while $I(J)\in\{x,y,z\}$ denote the various spin channels. By assuming a non-interacting ground state and taking the vertex to be the identity, $\varepsilon^{-1}_{IJ}$ is the RPA dielectric function and $P_{KL}$ the Lindhard polarizability. Since the electronic dispersion is local-density-approximation-like for x=0.12 \cite{Sahrakorpi2008} we use the single-band electronic dispersion $\varepsilon_{k}$ as defined in Ref. \onlinecite{Lane2020,Markiewicz2005} with the hopping parameters given in Table \ref{Table:hopping}. 

\begin{table}[h]
\centering
\caption{Tight-binding hopping parameters (in meV) for
La$_2$CuO$_4$ used in this study.}
\begin{tabular}{|c|c|c|c|}
$t$&$t^{\prime}$&$t^{\prime\prime}$&$t^{\prime\prime\prime}$\\
\hline
312.5 & -31.25 & 25.0 & 68.75 \\
\hline
\end{tabular}\label{Table:hopping}
\end{table}

To generate acoustic plasmons we must go beyond the usual Hubbard limiting case and consider the full long-range Coulomb interaction. Moreover, the treatment of the long-range Coulomb interaction for the correlated electronic system in a layered structure requires some care. We model $v(r)$ as $\sum_{i}\bar{v}(\mathbf{R}_i)\delta(r-\mathbf{R}_i)$, so that $v(\mathbf{q})=\sum_i \bar{v}(\mathbf{R}_i) \exp(-i\mathbf{q}\cdot\mathbf{R}_i)$. $\bar{v}(\mathbf{R}_i)$ is taken to be an on-site Hubbard $U=2$ eV [Ref. \onlinecite{Markiewicz2017}] for $\mathbf{R}_i =0$ and a screened Coulomb interaction contribution $\bar{v}(\mathbf{R}_i)=e^2/(\varepsilon_0 \mathbf{R}_i)$ for $\mathbf{R}_i \neq 0$, using a background dielectric constant $\varepsilon_0 = 6$. To recover the correct $q\rightarrow 0$ limit of $v(\mathbf{q})$ we range separate $v(\mathbf{r})$ into short-range (SR) and long-range (LR) pieces. For the short-range interactions where $r$ less than or equal to a cut-off $L$ we sum the contributions of all in-plane Cu terms. The remaining long-range component $(r>L)$ we approximate by a continuum. Then after summing over the interplane contribution we arrive at the following expression for $v(\mathbf{q})$  \cite{markiewicz2007, markiewicz2008},
\begin{align}\label{eq:vq}
v(\mathbf{q})=
v^{SR}_{2d}(\mathbf{q})+
v^{LR}_{2d}(\mathbf{q})+
v_{z}(\mathbf{q}),
\end{align} 
where
\begin{widetext}
\begin{subequations}
\begin{align}\
v^{SR}_{2d}(\mathbf{q}) &= U+\sum_{i\neq 0}^{L}\frac{e^2}{\varepsilon_{0}R_{i}}e^{-i\mathbf{q}\cdot \mathbf{R}_{i}},\\
v^{LR}_{2d}(\mathbf{q})&=\frac{2\pi e^2}{\varepsilon_{0}a^2 q_{\parallel}}\left[  1-q_{\parallel}LJ_{0}(q_{\parallel}L)\frac{\pi q_{\parallel}L}{2}\left( J_{0}(q_{\parallel}L)H_{1}(q_{\parallel}L)-J_{1}(q_{\parallel}L)H_{0}(q_{\parallel}L) \right)\right],\\
v_{z}(\mathbf{q})&=\frac{2\pi e^2}{\varepsilon_{0}a^2 q_{\parallel}}\left[ \frac{\cos(q_{z}l)-e^{-q_{\parallel}l}}{\cosh(q_{\parallel}l)- \cos(q_{z}l)} \right].
\end{align}
\end{subequations}
\end{widetext}
$a$ and $l=c/2$ are the in-plane lattice parameter and distance between adjacent CuO$_2$ planes in La$_2$CuO$_4$ \cite{Lane2018,Furness2018}, respectively, $J_i$ $(H_i)$ are the Bessel (Struve) functions, 
and the short-range cut-off $L$ is $500a$. Then by inserting $v(q)$ in Eq.~\ref{eq:dielectric} we obtain the loss function. 

\begin{figure*}[h!]
\centering
\includegraphics[width=\textwidth]{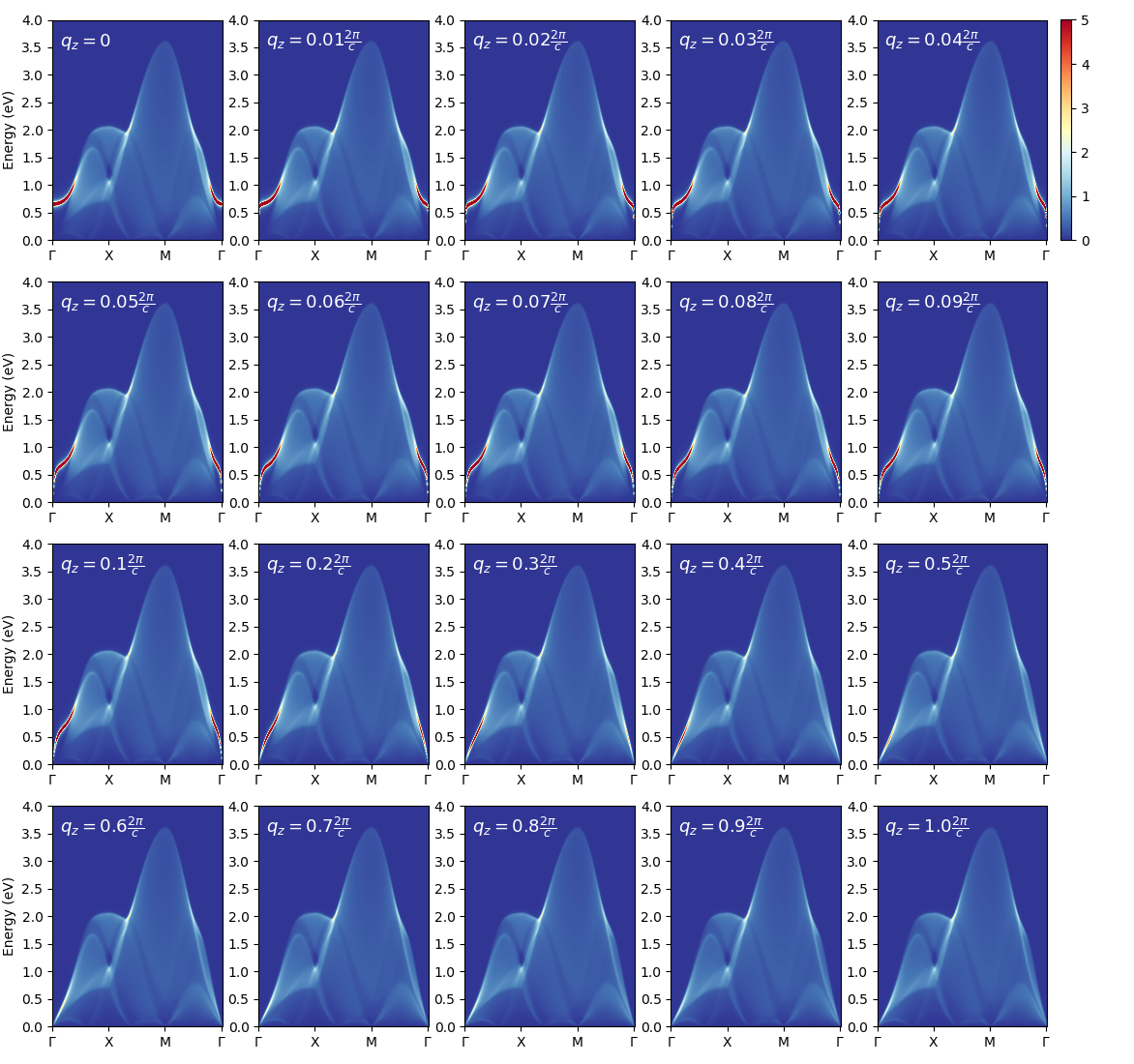}
\caption{The loss function along the high-symmetry lines in the square Brillouin zone for various values of $q_z$.}\label{fig:loss}
\end{figure*}

Figure~\ref{fig:loss} shows the loss function for values of $q_z$ ranging from $0$ to $1$ in units of $2\pi/c$. For $q_z=0$ [Fig~\ref{fig:loss} (upper-left panel)] a clear plasmon peak is seen at the zone center near 0.70 eV. As $q_\parallel$ changes from $\Gamma$ to $X$ the plasmon peak disperses to higher energies eventually entering the particle-hole continuum around $(\pi/2,0)$, where the peak intensity sharply decreases. As $q_z$ increases the plasmon energy at $\Gamma$ decreases, becoming zero for $q_z>0.04$. Moreover, the plasmon dispersion along $\Gamma-X$ softens becoming linear for $q_z>0.3$, ultimately merging with the electron-hole continuum for very large $q_z$, which dampens out the plasmon peak entirely. The evolution of the plasmon with $q_z$ is in agreement with Ref.~\onlinecite{kresin1988layer}.

\begin{figure*}[h!]
\centering
\includegraphics[width=\textwidth]{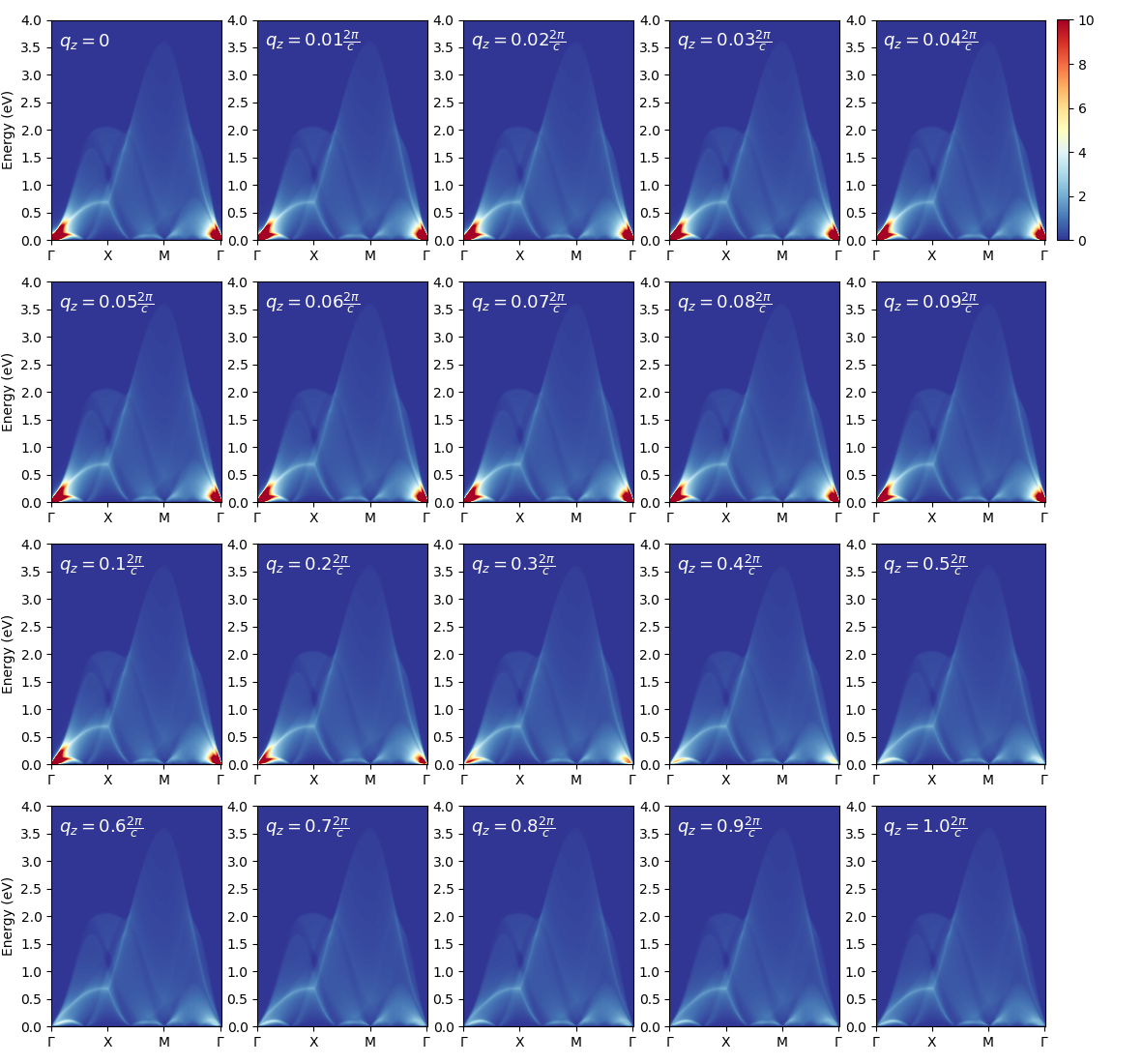}
\caption{The Imaginary part of $\varepsilon_{00}(\mathbf{q},\omega)$ along the high-symmetry lines in the square Brillouin zone for various values of $q_z$.}\label{fig:dielecIm}
\end{figure*}

\begin{figure*}[h!]
\centering
\includegraphics[width=\textwidth]{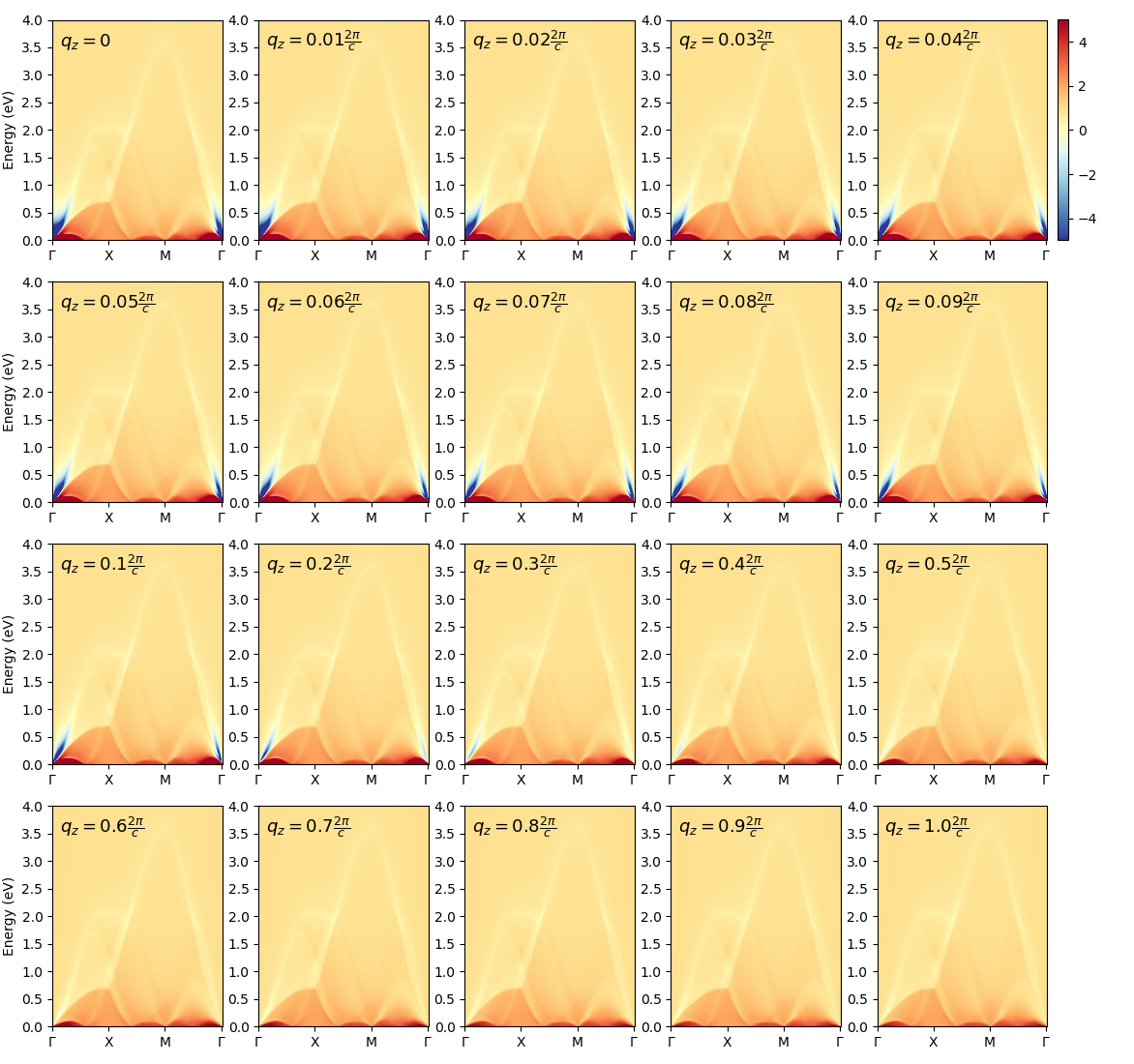}
\caption{The real part of $\varepsilon_{00}(\mathbf{q},\omega)$ along the high-symmetry lines in the square Brillouin zone for various values of $q_z$.}\label{fig:dielecRe}
\end{figure*}

Figure~\ref{fig:dielecIm} and~\ref{fig:dielecRe} present the real and imaginary parts of $\varepsilon_{00}(\mathbf{q},\omega)$ for $q_z$ varying from $0$ to $1$ in units of $2\pi/c$. When $q_z<<1.0$ a zero energy Drude peak appears in $\text{Im}\left[\varepsilon_{00}\right]$ due to the singularity in the bare Coulomb potential $v(\mathbf{q})$. However, as $q_z$ increases to $\sim 1.0$  the divergence in $v(\mathbf{q})$ at $q_\parallel=0$ is softened, producing a dielectric function which is approximately just the bare Lindhard polarization. Moreover, to fully capture the experimetnal dielectric function as reported by Uchida {\it et al.} \cite{uchida1991optical} the inclusion of self-energy effects are crutial\cite{das2010optical}, however these effects are not necessary to model the plasmon dispersion for this present study. 

Since the real part of the dielectric function is related to the imaginary part through the Hilbert transform, the energy dependence behaves as the derivative of the imaginary part. Therefore, zeros in $\text{Re}\left[\varepsilon_{00}(\mathbf{q},\omega)\right]$ are generated in response to peaks in $\text{Im}\left[\varepsilon_{00}\right]$ at special energies $\omega^*$. These zeros in the real part are seen in Fig.~\ref{fig:dielecRe} along $\Gamma-X$ at the boundary between the blue and red shaded regions. When $q_z$ is small $\text{Re}\left[\varepsilon_{00}\right]$  exhibits a sharp decrease to negative values gradually becoming positive again around 0.75 eV. At this node both $\text{Im}\left[\varepsilon_{00}\right]$ and $\text{Re}\left[\varepsilon_{00}\right]$ are zero, thus generating a plasmon pole when the dielectric function is inverted. The evolution of this pole with $q_z$ follows the plasmon dispersion in Fig.~\ref{fig:loss}.

\begin{figure*}[h!]
\centering
\includegraphics[width=17.5cm]{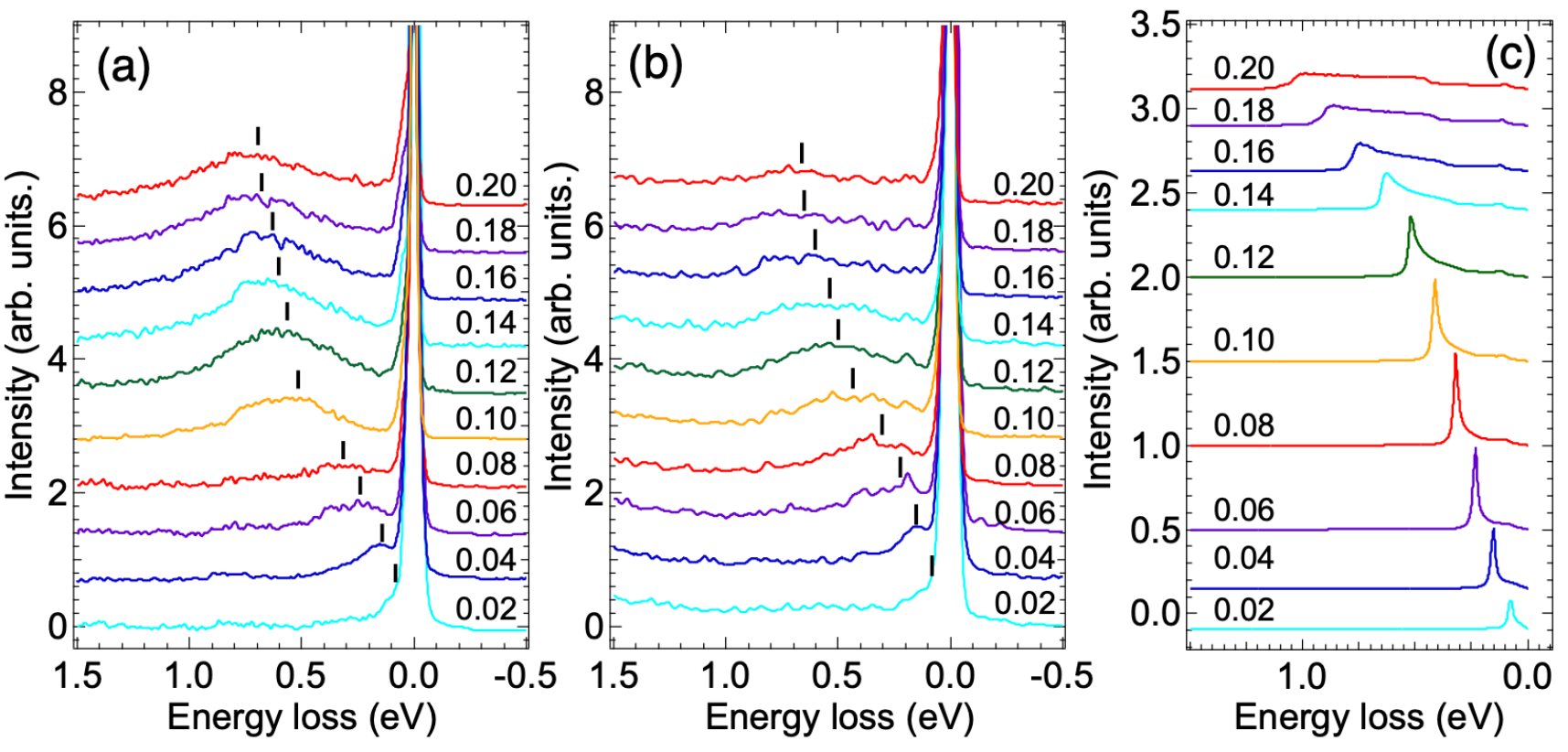}
\caption{ (a) \& (b) Momentum-dependent RIXS spectra of LSCO excited by X-rays of energy tuned to the ZRS and the UHB resonances, respectively. The out-of-plane component $q_{z}$ was fixed to 0.7.  Colour curves in (a) \& (b) are the same as those of Fig. 3 in the main text.  (c) Calculated loss functions for $q_{z} = 0.7$ extracted from Fig. \ref{fig:loss}. All spectra are plotted with $q_\|$ indicated and offset vertically for clarity. Vertical ticks in (a) and (b) indicate the plasmon energies.}\label{fig_s7}
\end{figure*}

Figure~\ref{fig_s7} compares the calculated loss functions with the measured RIXS spectra for various $q_{\|}$. Our calculations were performed at nearly 0 K and did not include any lifetime effects for the acoustic plasmons. In addition, some other competing excitations that may smear out the plasmon feature for $q_z$ approaching 0, especially since the Coulomb potential is becoming more long range and therefore could be driving other excitation modes. For example, an exciton may appear at similar energy scales \cite{Collart2006}. Therefore, the line shape of the calculated loss function is much sharper than that of RIXS excitations of a real material.

\clearpage
\bibliographystyle{naturemag}
\bibliography{supplement.bib}